\documentclass[journal]{IEEEtran}
\usepackage{cite,calc}
\usepackage[cmex10]{amsmath}   
\usepackage{amsthm}
\usepackage{amsfonts,amssymb,graphics,psfrag,subfigure,epsfig,graphics}
\usepackage[mathcal]{euscript}  
\usepackage{setspace}
\usepackage{color}

\newtheorem{Theorem}{Theorem}

\newtheorem{Lemma}[Theorem]{Lemma}

\newtheorem{Example}{Example}

\newcommand{\nchoosek}[2]{\binom{#1}{#2}}

\ifCLASSINFOpdf
\else
\fi
\begin{document}
%
\title{Network Information Theoretic Security} 
%
%
%

\author{Hongchao~Zhou
        and~Abbas~El Gamal
\thanks{H. Zhou is with the School of Information Science and Engineering, Shandong University, Qingdao,
Shandong, 266237 China.}
\thanks{A. El Gamal is with the Department of Electrical Engineering,
Stanford University, Stanford, CA,  94305 USA.}}
\maketitle

\begin{abstract}
Shannon showed that to achieve perfect secrecy in point-to-point communication, the message rate cannot exceed the shared secret key rate giving rise to the simple one-time pad encryption scheme. In this paper, we extend this work from point-to-point to networks. We consider a connected network with pairwise communication between the nodes. We assume that each node is provided with a certain amount of secret bits before communication commences. An eavesdropper with unlimited computing power has access to all communication and can hack a subset of the nodes not known to the rest of the nodes. We investigate the limits on information-theoretic secure communication for this network. We establish a tradeoff between the secure channel rate (for a node pair) and the secure network rate (sum over all node pair rates) and show that perfect secrecy can be achieved if and only if the sum rate of any subset of unhacked channels does not exceed the shared unhacked-secret-bit rate of these channels.
We also propose two practical and efficient schemes that achieve a good balance of network and channel rates with perfect secrecy guarantee.  This work has a wide range of potential applications for which perfect secrecy is desired, such as cyber-physical systems, distributed-control systems, and ad-hoc networks.
\end{abstract}


%
\IEEEpeerreviewmaketitle

\section{Introduction}

The information-theoretic security introduced by Shannon~\cite{Shannon1949} and widely accepted as the strictest notation of security, is becoming increasingly attractive for many cyber-physical systems, distributed-control systems, wireless ad-hoc networks, among other applications. Secure network coding~\cite{Cai2011} has been well studied to guarantee the information-theoretic security when a subset of channels are wiretapped~\cite{Cui2012,Cai2011a} or in the presence of Byzantine adversaries~\cite{Ho2004,Jaggi2007}. In this paper, we make a stronger assumption: all the channels are eavesdropped and some nodes are hacked without knowledge of the rest of the nodes. This assumption is realistic for example in wireless networks in which an eavesdropper can sense the transmitted signals, or the network nodes communicate via an insecure public network as another example. Under this assumption, pure network-coding approaches cannot work without the help of common randomness shared among the network nodes.
Physical layer security~\cite{Bloch2008, Shiu2011, Liang2009, Liu2017, Dong2010} can be used to distribute secret keys among network nodes, however, the channel advantage required by the receivers over any eavesdropper is not easy to guarantee in a wireless network.
In many scenarios, it is feasible and much cheaper to pre-distribute a very large number of secret bits to network nodes to support future secure communication. In this paper, we are interested in a fundamental problem: if every node in the network is allowed to carry a certain large number of secret bits, how much information can be securely transmitted over the network under the information-theoretic security criterion?

We consider a connected network of $n$ nodes, with at most $t\leq n-2$ nodes hacked without knowledge of the other nodes. We assume pairwise communication with end-to-end encryption, that is each sender node encrypts its message using a secret key generated from the common randomness shared with the intended receiver node and the receiver decrypts it using the same key. Through the process of key pre-distribution, each node has a sequence of $l$ secret bits, and the secret bits from different nodes may be correlated according to a symmetric joint probability distribution which does not depend on future communications. If a node is hacked, all its secret bits are disclosed to the eavesdropper. Before two nodes communicate, they generate a secret key by purifying their common randomness with privacy amplification~\cite{Bennett1988,Maurer1993,Maurer2000,Bennett1995} then use the one-time pad scheme to communicate the message. As the total length of the messages to be transmitted over a channel (i.e., from a source to a destination) cannot exceed the secret-bit length $l$ of each node, we define their ratio as the channel rate, and the sum rate of all the channels as the network rate.

The secure-communication limit of a network depends on the secret bits distributed and the method of utilizing them to deliver information. As an example, consider a network with $n=4$ nodes and $t=1$ nodes being hacked. A straightforward way to pre-distribute the secret bits is to assign each pair of nodes $l/3$ common secret bits as the secret key which they can use with the one-time pad scheme.  In this case, the messages are secure if and only if the rate of each channel does not exceed $1/3$. As the network size $n$ increases, this approach can only reach channel rate of at most $1/(n-1)$, limiting its applications for large networks. An alternative way to redistribute the secret bits in the $4$-node example is to assign every three nodes $l/3$ common secret bits, and hence there are four sequences of secret bits denoted by $\mathbf{u}_\textrm{123}, \mathbf{u}_\textrm{124}, \ldots$, where $\mathbf{u}_{123}$ is the secret bits distributed to nodes $1,2,3$. When two nodes say nodes $1$ and $2$ communicate to each other, they use $\mathbf{u}_\textrm{123} + \mathbf{u}_\textrm{124}$ as the secret key. In this case, no matter whether node $3$ or node $4$ is hacked, the messages are secure as long as the channel rate between node $1$ and $2$ does not exceed $1/3$. Compared to the previous way, it can be proved that for a larger network of size $n$, by allowing each secret bit to be distributed to multiple nodes instead of only two nodes, the maximum channel rate (channel capacity) can be improved to more than $1/4$ from $1/(n-1)$. Our scheme can be viewed as another application of linear network coding. It allows higher secure communication rates by utilizing the secret bits shared by multiple channels perfect secrecy can be achieved.

We address several basic questions about our network setting: (i) what is the limit on the network rate and the channel rate for secure communication considering all the symmetric ways of pre-distributing secret bits? (ii) given an arbitrary distribution on the secret bits, how do we determine the security of a network with given channel rates? and (iii) how to design efficient and practical network-communication schemes that have both high network rate and high channel rate?

The rest of this paper is organized as follows. Section \ref{section_limit} provides the formulation and definitions of the problem for network communication with information theoretic security. Section \ref{section_summary} summaries some main results. Two practical and efficient communication schemes with perfect secrecy guarantee are proposed and
investigated in Section \ref{section_schemes}. Section \ref{section_discussions} further discusses the network security beyond the information theoretic limit, the application of network coding, and some open questions. The proofs of the main results are given in Section \ref{section_proofs}.

\section{Definitions}
\label{section_limit}

We consider a network consisting of a set of nodes $\mathcal{N}=\{1, 2, \ldots, n\}$, where every two nodes can find a path (channel) connecting them, and use $\mathcal{P} = \{(i, j)\}$ to denote the set of all the channels. It is assumed that all the channels are insecure, namely, every ciphertext transmitted over the network is possible to be known by an eavesdropper. Each node in the network is able to store a large number $l$ of secret bits. To guarantee the network security, the total message length $m_{ij}$ of a channel $(i,j)\in \mathcal{P}$ cannot exceed $l$. We call the number of message bits transmitted through a channel $(i,j)$ per a node's secret bit as the \emph{channel rate} $r_{ij}$, and the sum of the channel rates as the \emph{network rate} $r$. Mathematically,
$$r_{ij} = \frac{m_{ij}}{l}, \quad r=\sum_{(i,j)\in \mathcal{P}}r_{ij}.$$

We assume that up to $t$ nodes could be hacked by an eavesdropper, and in this case, all the secret bits stored in these nodes are revealed to the eavesdropper. We use $\mathcal{N}_\textrm{h} \subset \mathcal{N}$ with $|\mathcal{N}_\textrm{h}|\leq t$ to denote the set of hacked nodes, and use $\mathcal{N}_\textrm{s} = \mathcal{N}/\mathcal{N}_\textrm{h}$ to denote the set of unhacked secure nodes. As a channel is insecure if one of its terminals is hacked, our goal is to protect those channels between secure nodes, denoted by $\mathcal{P}_\textrm{s}$. Note that if $t\geq n-1$, it has $|\mathcal{P}_\textrm{s}|=0$, and in this case no message bits can be securely transmitted between any two nodes. In this paper, we assume that $t\leq n-2$ by default.

A secure network-communication scheme $\psi$ consists of two phases. In the key pre-distribution phase, the scheme provides a sequence of secret bits $\mathbf{u}\in \{0, 1\}^*$ to the network nodes, with each bit possibly distributed to multiple nodes. We use $\mathbf{u}_i$ to denote the sequence of secret bits distributed to node $i$. In the communication phase, we let $\mathbf{x}_{ij}\in \{0,1\}^{m_{ij}}$ be the message transmitted in channel $(i,j)$ (we assume there is a single message transmitted in each channel for simplicity), and the corresponding ciphertext is $\mathbf{y}_{ij}$. Both all the transmitted ciphertexts $\mathbf{y} = \{\mathbf{y}_{ij}|(i,j)\in \mathcal{P}\}$ and the hacked secret bits $\mathbf{u}_\textrm{h}= \cup _{j\in \mathcal{N}_\textrm{h}} \mathbf{u}_j$ are known to the eavesdropper.

The network communication is information-theoretically secure if and only if for any possible set of hacked nodes $\mathcal{N}_\textrm{h}$ and for any channel
$(i, j) \in \mathcal{P}_\textrm{s}$, it has
\begin{equation}
\textrm{I}[\mathbf{x}_{ij}; \mathbf{y}, \mathbf{u}_\textrm{h}] < \epsilon
\end{equation}
for small $\epsilon \ge 0$, with $\epsilon=0$ for perfect secrecy.  Note that the security of the network depends on the message lengths $\{m_{ij}\}$ (i.e. the channel rates $\{r_{ij}\}$ for a given $l$) but not on their exact values or which of the two terminals is the source. We say that a set of channel rates $\{r_{ij}\}$ is achievable using a scheme $\psi$ if any messages with these channel rates can be securely communicated with high probability (arbitrarily close to $1$) for $l$ sufficiently large. We denote the set of achievable rates using a scheme $\psi$ by $\mathcal{R}(\psi)$.

We define the \emph{maximum channel rate} and the \emph{maximum network rate} of a scheme $\psi$ as the supremum of all the achievable channel rates and network rates of the scheme, and denote them by
$$\emph{R}_\textrm{channel}(\psi)=\sup_{\{r_{ij}\}\in \mathcal{R}(\psi)} \max_{(i,j)\in \mathcal{P}} r_{ij},$$
$$\emph{R}_\textrm{net}(\psi)=\sup_{\{r_{ij}\}\in \mathcal{R}(\psi)} \sum_{(i,j)\in \mathcal{P}} r_{ij}.$$

The communication demand is typically unknown during the key pre-distribution phase. To ensure that the designed schemes have enough flexibility, the secret bits are pre-distributed to network nodes in a symmetric fashion (thus permuting the indices of the nodes  does not change the joint probability distribution of the secret bits). In this case, the maximum network rate can be achieved by the equal channel rates with $r_{ij} = \frac{1}{n-1}$ for all the channels, following the symmetry assumption of the network as well as the distributed randomness. On the other hand, the maximum channel rate can be achieved when there is only one channel having message transmissions, i.e., $r_{ij}>0$ for a specific single channel and $r_{ij}= 0$ for all the other channels.

To investigate the communication limit, we define the \emph{channel capacity} and the \emph{network capacity} of a network as the maximum over all the maximum channel rates and maximum network rates of any scheme with symmetric key distribution, and denote them by
$$\emph{C}_{\textrm{channel}} = \max_\psi \emph{R}_{\textrm{channel}} (\psi), \quad \emph{C}_{\textrm{net}} = \max_\psi \emph{R}_{\textrm{net}} (\psi).$$

\section{Summary of Main Results}
\label{section_summary}

From its definition, the maximum channel rate of a scheme is at most $1$. But this upper bound cannot be reached when $t>0$. The following result provides the capacities of
a general network, as the theoretical limit for all the schemes with symmetric key distribution.
From this result, a network with $n=4$ and $t=1$ has channel capacity of $1/3$. Interestingly, if we further increase the size of the network to $5$, the channel capacity is still $1/3$.

\begin{Theorem}\label{theorem2}
Given a network of $n$ nodes with at most $t\leq n-2$ nodes being hacked, its network capacity and channel capacity are
$$\emph{C}_{\textrm{net}}=\frac{n}{2},\quad \emph{C}_{\textrm{channel}}= \frac{\nchoosek{n-t-2}{a-2}}{\nchoosek{n-1}{a-1}} \textrm{ with } a= \lceil\frac{n}{t+1}\rceil.$$
\end{Theorem}

There is a certain tradeoff between the maximum network rate and the maximum channel rate of a scheme. In particular, they have the following relationship for $t=0$, implying that one may sacrifice the maximum network rate to obtain a better maximum channel rate, and vice versa.

\begin{Theorem} \label{theorem1}Given a network of $n$ nodes without any nodes being hacked, for any scheme $\psi$ with symmetric key distribution, it satisfies
\begin{equation}\emph{R}_\textrm{net}(\psi) \frac{2}{n+1} + \emph{R}_\textrm{channel} (\psi) \frac{n-1}{n+1} \leq 1. \label{equ3_12}\end{equation}
The equality is achievable for any $1\leq \emph{R}_\textrm{net}(\psi) \leq \frac{n}{2}$.
\end{Theorem}

Given a scheme, it is crucial to determine whether a network with channel rates $\{r_{ij}\}$ is perfectly secure or not, as the communication phase has to be terminated for guaranteeing perfect secrecy when the channel rates get very close to the limit. The difficulty arises from the fact that different channels may share some common secret bits, hence ``interfere" with each other. Given an arbitrary (symmetric or not) distribution of secret bits, we prove that perfect secrecy is achievable if and only if the sum rate of any subset of unhacked channels does not exceed the shared unhacked-secret-bit rate of these channels. Here, given
the set of hacked nodes $\mathcal{N}_\textrm{h}$, the shared unhacked-secret-bit rate of a set of channels $P \subseteq \mathcal{P}_\textrm{s}$ is defined by
$$r_\textrm{secrecy}(P, \mathcal{N}_\textrm{h}) = \frac{|\cup_{(i,j)\in P} \mathbf{u}_{ij} / \mathbf{u}_\textrm{h}|}{l}.$$

\begin{Theorem}\label{theorem4}
Given a network of $n$ nodes with at most $t\leq n-2$ nodes being hacked, the channel rates $\{r_{ij}\}$ are achievable if and only if for any possible set of hacked nodes $\mathcal{N}_\textrm{h}$ and for any subset of channels $P \subseteq \mathcal{P}_\textrm{s}$, it satisfies \footnote{It is assumed that privacy amplification is applied to every channel. Otherwise, the equality holds if the common secret bits shared by two nodes are unknown by the rest of nodes and are used as the secret key directly.} either $\sum_{(i,j)\in P} r_{ij}=0$ or
\begin{equation}\sum_{(i,j)\in P} r_{ij}< r_\textrm{secrecy}(P, \mathcal{N}_\textrm{h}).\label{equ3_3}\end{equation}
\end{Theorem}

Achievability uses a simple method for privacy amplification that generates each secret-key bit by computing the XOR of $d$ randomly sampled common secret bits shared by two terminals with $d=O(\log l)$.
For a network with $4$ nodes, assume that the secret bits are distributed as follows: each secret bit is distributed to $3$ different nodes, and every $3$ nodes share $l/3$ common secret bits. If no nodes are hacked, then the network is secure for sufficiently large $l$ if and only if
\begin{align*}
    r_{12} & < 2/3\\
     r_{12} + r_{13} + r_{14} & < 1\\
    \sum_{ij} r_{ij} & < 4/3
\end{align*}
holds for any node permutation. None of the inequalities can be dismissed. If one of the nodes is hacked, then the network is secure for sufficiently large $l$ if and only if \begin{equation}r_{12} + r_{13} + r_{23} < 1/3 \label{equ1_2}\end{equation}
holds for any node permutation.

Note that if the size of the network is large, the number of constrains in the above criteria becomes prohibitively large. One can reduce the computational complexity by relaxing the conditions to hold only for any set of channels of some size $w$,
$$\max_{|P|=w}\sum_{(i,j)\in P} r_{ij} < \min_{|P|=w, \mathcal{N}_\textrm{h}} r_\textrm{secrecy}(P, \mathcal{N}_\textrm{h}),$$
where the left term is easy to calculate and the right term can be computed explicitly (see subsection \ref{relaxed_criteria}) for symmetric key distribution.

The following result provides an alternative approach to check the security of a network, in which we let $\mathbf{u}_G$ be the set of secret bits distributed only to all the nodes in set $G$. The shared unhacked-secret-bit rate of $G$ is defined by $$r_\textrm{secrecy}(G, \mathcal{N}_h) = \frac{|\mathbf{u}_G/ \mathbf{u}_\textrm{h}|}{l}.$$

\begin{Theorem}\label{theorem5}
Given a network of $n$ nodes with at most $t\leq n-2$ nodes being hacked, the channel rates $\{r_{ij}\}$ are achievable if for any possible set of hacked nodes $\mathcal{N}_\textrm{h}$, there exists a  non-negative feasible solution for $\{x_{G}^{ij}\}$ such that
\begin{align*}
& r_{ij} < \sum_{G| i,j\in G \subseteq \mathcal{N}_\textrm{s}} x_{ij}^{G}, \quad \forall (i, j) \in \mathcal{P}_\textrm{s} \textrm{ }\textrm{with} \textrm{ }r_{ij} > 0 ,\\
    & r_\textrm{secrecy}(G, \mathcal{N}_h)  = \sum_{i,j\in G} x_{ij}^{G}, \quad \forall G \subseteq \mathcal{N}_\textrm{s}.
\end{align*}
\end{Theorem}

We continue using the network of $4$ nodes as an example. If node $4$ is hacked, then only $\mathbf{u}_{123}$ is not hacked. The network is secure for sufficiently large $l$ if there exists a feasible solution for $\{x_{12}, x_{13}, x_{23}\}$ such that
$$r_{12} < x_{12}, r_{13} < x_{13}, r_{23} < x_{23},$$
 $$1/3  = x_{12} + x_{13} + x_{23},$$
reaching the same condition as (\ref{equ1_2}).

\section{Network Schemes}
\label{section_schemes}

We wish to develop schemes that can securely communicate as many message bits as possible not only over the entire network but also through a single channel.

As defined earlier, a secure network-communication scheme consists of a key pre-distribution phase and a communication phase in which a secret key between two nodes is established from their common secret bits via privacy amplification and the one-time pad scheme is used to achieve secure communication. There are a variety of methods for privacy amplification, such as universal hashing~\cite{Carter1979}, random linear transformations~\cite{Zhou2013} and polar codes~\cite{Chou2015}. Given a large number of common secret bits $\mathbf{u}_{ij}$ between two nodes, one straightforward idea is to divide the shared secret bits into blocks, and then apply privacy amplification to each block. However, this approach is not appropriate for our applications as it requires knowledge of the message length (channel rate) before communication as well as sophisticated coordination among the nodes. We adopt a simple method for privacy amplification: each secret-key bit is generated by computing the XOR of $d$ randomly sampled common secret bits from $\mathbf{u}_{ij}$ with $d$ a large integer, for example, $128$.  We can repeat this process whenever more secret-key bits are needed. The performance of this method is very close to optimal. To further improve the computational efficiency, one can generate $q\gg 1$ secret-key bits simultaneously by packing $q$ secret bits together at the same location and performing the same operations on them.

\subsection{Key Pre-Distribution}

We study two different key pre-distribution schemes. The first is the combinational key scheme, which distributes each secret bit to exactly $a\geq 2$ nodes. The second scheme is the random key scheme, which distributes each secret bit to every node with some probability $p\leq 1$.

The combinational key scheme distributes the same number of distinct secret bits to each combination of $a$ nodes with $a\geq 2$. Hence, we divide all the secret bits into $\nchoosek{n}{a}$ groups each of size $l/\nchoosek{n-1}{a-1}$, and assign the secret bits of each group to a unique combination of $a$ nodes. For every two nodes, their shared common secret bits consist of the secret bits from $\nchoosek{n-2}{a-2}$ groups. There is a problem with this scheme: when $n$ and $a$ are large, the scheme becomes less practical as there are too many groups of secret bits. To address this problem, we suggest to use only  $m$ random groups. This subset of $m$ groups can be found based on an $m\times n$ random matrix with each row containing $a$ ones (corresponding to a group) and each column containing $am/n$ ones (corresponding to a node), whose construction has been studied for the parity-check matrix of regular LDPC codes~\cite{Gallager1963}.

The random key scheme generates a random-bit sequence $\mathbf{u}$ of length $u$ and assigns each of its bits to every network node with a predetermined probability $p\leq 1$.
Similar ideas were explored for key management in sensor networks with the computational security~\cite{Eschenauer2002, Chan2003}. In contrast, we study the key distribution for the information theoretic security, which directly affect the communication rates. With the random key scheme, each node $i$ obtains a sequence of secret bits $\mathbf{u}_i$ with length around $pu\simeq l$. A difficulty with this scheme is to help every two nodes to identify their shared common secret bits for generating a secret key. It would be too storage-inefficient if each node stores not only the values of its secret bits but also their locations in $\mathbf{u}$.
Our observation is that the secret bits need to be truly random, but not the ways of distributing them. One can use a pseudo-random permutation for key pre-distribution, and it helps to identify those common secret bits between nodes. Specifically, given the network size $n$ and the total number of secret bits $u$, we construct a pseudo-random permutation~\cite{Katz2007} $$\textrm{F}: \{1,2, \ldots, u\}\times \{1,2,\ldots,n\}\rightarrow  \{1,2, \ldots, u\}.$$ We distribute the $k$th secret bit in $\mathbf{u}$ to node $i$ at the location $F(k,i)$ if $F(k,i)\leq l$. This pseudo-random permutation is publicly known by all the network nodes.

\subsection{Maximum Rates}

\begin{table*}
  \centering
  \begin{spacing}{1.5}
  \begin{tabular}{|c|c|c|}
    \hline
      & Maximum Network Rate & \quad\quad Maximum Channel Rate \quad\quad\\
    \hline
    Combinational Key Scheme &  $\frac{n}{a}\frac{\nchoosek{n-t-2}{a-2}}{\nchoosek{n-2}{a-2}}$ & $\frac{a-1}{n-1}\frac{\nchoosek{n-t-2}{a-2}}{\nchoosek{n-2}{a-2}}$  \\
    \hline
    Random Key Scheme & $\frac{1}{p}\frac{\nchoosek{n}{2}}{\nchoosek{n-t}{2}}((1-p)^t - (1-p)^{n}- (n-t) p(1-p)^{n-1})$ &  $p (1-p)^t$\\
    \hline
  \end{tabular}
  \end{spacing}
  \caption{The maximum rate of the combinational key scheme and the random key scheme.}\label{table_capacities}
\end{table*}

Table \ref{table_capacities} lists the maximum rates of the combinational key scheme and the random key scheme. For both the schemes, the maximum network rates can be achieved by the equal channel rates with $r_{ij}=\frac{1}{n}$ and the maximum channel rates can be achieved when there is only one channel with message transmissions. Let us take the combinational key scheme as an example.
\begin{Example} For the combinational key scheme with $a\geq 2$, each node is distributed $\nchoosek{n-1}{a-1}$ groups of secret bits, and every two nodes share
$\nchoosek{n-2}{a-2}$ groups of secret bits. When $t$ nodes are hacked, then for every pair of unhacked nodes, they share $\nchoosek{n-t-2}{a-2}$ groups of secret bits that are not hacked. As a result, the maximum channel rate of the combinational key scheme is
$$\emph{R}_\textrm{channel}(\psi_\textrm{comb}) = \nchoosek{n-t-2}{a-2} / \nchoosek{n-1}{a-1} = \frac{a-1}{n-1}\frac{\nchoosek{n-t-2}{a-2}}{\nchoosek{n-2}{a-2}}.$$
\end{Example}

For the maximum rates of the combinational key scheme, they have a common term $\gamma(t,a) = \nchoosek{n-t-2}{a-2}/\nchoosek{n-2}{a-2}$ which is a decreasing function of $t$ with $\gamma(0,a)=1$. This term captures the effect of the number of hacked nodes $t$ on the maximum rates of the scheme.
From this term, we can estimate the number of hacked nodes that the scheme can tolerate. For instance, when $a=3$,  $\gamma(t,a) = \frac{n-t-2}{n-2}$, and one can tolerate relatively large $t$. When $a$ is large, the scheme can only tolerate a very small number of nodes to be hacked. We observe similar behaviors for the maximum rates of the random key scheme, which have a common term approximately $(1-p)^t$ that captures the effect of $t$.

Further comparing the maximum rates of the two schemes, there is a rough mapping between the parameter $a$ in the combinational key scheme and the $np$ in the random key scheme, which is about the expected number of nodes that each secret bit is distributed to. The intuition behind this mapping is that: compared to the scheme that distributes each secret bit to $2$ nodes, the proposed schemes distribute each secret bit to around $a\ge 2$ nodes. As a result, the maximum channel rates of the schemes increase by a factor of $a-1$. Meanwhile, the usage efficiency of each secret bit (corresponding to the maximum network rates) is reduced by a factor of roughly $\frac{2}{a}$, as each secret bit can only be used for once by two nodes among the $a$ nodes.

\subsection{Hybrid Schemes}

For any combinational/random key scheme, the multiplication of its maximum network rate and its maximum channel rate does not exceed $1$.
It implies that with a pure combinational/random key scheme, high network rate and high channel rate cannot be achieved at the same time.

We denote the combinational key scheme with $a=2$ as the pairwise key scheme, which reaches the network capacity. To better balance the maximum network rate and the maximum channel rate, we consider a hybrid scheme, in which each node uses a fraction $\lambda \in [0,1]$ of its storage space to run the pairwise key scheme and the rest to run the combinational key scheme with $a>2$ (or a random key scheme with $np>2$). The maximum network rate of this hybrid scheme is the weighted sum of their respective component schemes' maximum network rates, and so is its maximum channel rate. For $\lambda = \frac{1}{2}$, for example, the maximum rates of the hybrid scheme are
$$\emph{R}_\textrm{net}(\psi_\textrm{hybrid}) = \frac{n}{4} + \frac{n}{2a}\frac{\nchoosek{n-t-2}{a-2}}{\nchoosek{n-2}{a-2}},$$
$$\emph{R}_\textrm{channel}(\psi_\textrm{hybrid}) = \frac{1}{2(n-1)} + \frac{a-1}{2(n-1)}\frac{\nchoosek{n-t-2}{a-2}}{\nchoosek{n-2}{a-2}}.$$
The maximum network rate is strictly larger than $n/4$, and the maximum channel rate can be adjusted by selecting appropriate $a$. For example, for a network with $n=100$ nodes and $t=1$, the maximum network and channel rates of the pairwise key scheme are $50.0$ and $0.0101$, respectively, and those of the hybrid scheme with $a=25$ are $26.53$ and $0.0978$, respectively, which improves on the maximum channel rate by sacrificing on the maximum network rate.

\subsection{Security Criteria}
\label{relaxed_criteria}
When the network size is large, it is computationally too complex to check the security of a network directly based on Theorem \ref{theorem4} and Theorem \ref{theorem5},
as the number of  constrains in the criteria becomes prohibitively large. We discuss some techniques to reduce the number of constrains by relaxing the criteria, and their applications to the proposed schemes.

We can relax the conditions in the criteria of Theorem \ref{theorem4} to hold only for any set of channels of some size $w$,
$$\max_{|P|=w}\sum_{(i,j)\in P} r_{ij} < \min_{|P|=w, \mathcal{N}_\textrm{h}} r_\textrm{secrecy}(P, \mathcal{N}_\textrm{h}).$$
The left term is easy to calculate, and we would like find a simple way to compute the right term
$$r_\textrm{secrecy}(w) = \min_{|P|=w, \mathcal{N}_\textrm{h}} \frac{|\cup_{(i,j)\in P} \mathbf{u}_{ij} / \mathbf{u}_\textrm{h}|}{l}.$$

Due to the symmetry of key distribution, we can assume without loss of generality that the last $t$ nodes are hacked, and the minimum of $|\cup_{(i,j)\in P} \mathbf{u}_{ij}/ \mathbf{u}_\mathrm{h}|$ with $|P|=w$ can be reached by the first $w$ elements in
$$\mathbf{u}_{12}, \mathbf{u}_{13}, \ldots, \mathbf{u}_{1(n-t)}, \mathbf{u}_{21}, \ldots, \mathbf{u}_{(n-t-1)(n-t)},$$
whose indices are in lexicographical order. As a consequence, $r_\textrm{secrecy}(w)$ can be computed in an explicit way
\begin{equation}r_\textrm{secrecy}(w) = \frac{|(\mathbf{u}_{12}\bigcup\mathbf{u}_{13}\ldots\bigcup\mathbf{u}_{xy})/\mathbf{u}_\mathrm{h}|}{l}\end{equation}
for some $x,y$ with $1\leq x < y \leq n-t$ and
$$w=\nchoosek{n-t}{2} - \nchoosek{n-t-x}{2}-(n-t-y).$$
The number of constrains is therefore reduced to $\nchoosek{n-t}{2}$.

If we apply the relaxed criteria to the combinational key scheme with $a>2$, it can be shown that
$$r_\textrm{secrecy}(w)= \frac{\nchoosek{n-t}{a} - \nchoosek{n-t-x}{a} - \nchoosek{n-t-y}{a-1}}{\nchoosek{n-1}{a-1}}.$$

If we apply the relaxed criteria to the random key scheme with probability $p$, it can shown that
\begin{align*}
  r_\textrm{secrecy}(w) =& (1-p)^t[\alpha(n-t)/p - (1-p)^x\alpha(n-t-x)/p \\
  &- (1-p)^{y-1} (1-(1-p)^{n-t-y}) ],
\end{align*}
where $\alpha(n)= 1- (1-p)^n - n p (1-p)^{n-1}$ is the probability that a secret bit is distributed to at least two nodes among $n$ nodes.

For the criteria of Theorem \ref{theorem5}, instead of determining whether there exists a feasible solution of $\{x_G^{ij}\}$ for all possibilities of hacked nodes $\mathcal{N}_\mathrm{h}$, it is much easier to check whether a concrete solution is feasible. Specifically, given a set of hacked nodes $\mathcal{N}_\mathrm{h}$, we can construct $\{x_G^{ij}\}$ such that
\begin{equation}x_G^{ij} = \frac{\frac{|\mathbf{u}_G /\mathbf{u}_\mathrm{h}|}{|G|}}{\sum_{G': i,j \in G'\subseteq \mathcal{N}_\mathrm{s}} \frac{ |\mathbf{u}_{G'} /\mathbf{u}_\mathrm{h}|}{|G'|}}(1+\epsilon)r_{ij},\end{equation}
where $|G|$ is the number of nodes in $G$ and $\epsilon>0$ is small. This $\{x_G^{ij}\}$ satisfies $$r_{ij} < \sum_{G| i,j\in G \subseteq \mathcal{N}_\mathrm{s}} x_{G}^{ij}=(1+\epsilon) r_{ij}$$
for all $(i,j)\in \mathcal{P}_\textrm{s}$ with $r_{ij} >0$. If $$\sum_{i,j\in G} x_{G}^{ij} \leq r_\textrm{secrecy}(G, \mathcal{N}_\textrm{h})$$
holds for all $G \subseteq \mathcal{N}_\mathrm{s}$, then there must be a feasible solution such that the equalities hold, and the channel rates are achievable.
This can be used to check the security of the combinational key scheme.

\section{Discussions}
\label{section_discussions}

In this section we provide some further discussions, including the security strength with channel rates near their theoretical limit, the application of network coding, and several open questions.

\subsection{Security Beyond Limit}

Can a network continue to communicate when its channel rates reach or even exceed the theoretical limit? We demonstrate that the network communication near the theoretical limit is still more secure than cryptographic approaches that are based on some unproven assumptions about computational hardness.

Let us study a simplified model: let $\mathbf{u}\in \{0, 1\}^u$ be a sequence of random bits with $u$ very large, and a secret key $\mathbf{s}\in \{0, 1\}^m$ with $u < m < 2u$ is generated with a random linear transformation on $\mathbf{u}$, i.e.,
$\mathbf{s} = M \mathbf{u}$, where $M$ is a random matrix. Let $\mathbf{x}\in \{0, 1\}^m$ be the transmitted message, then the ciphertext generated using the one-time pad scheme is
$$\mathbf{y} = \mathbf{s} + \mathbf{x} = M\mathbf{u} + \mathbf{x}.$$
Attacking the system is to derive some information about $\mathbf{x}$ from $\mathbf{y}$ and $M$. Although perfect secrecy is not guaranteed, it is extremely difficult to derive some information about $\mathbf{x}$ when the dimension $u$ is very large for the following reasons. Firstly, the attacking process is analog to the decoding of a linear random code, with $\mathbf{u}$ being the message, $\mathbf{s}$ as the codeword, and $\mathbf{x}$ as the noise.
It has been proven that finding the $\mathbf{x}$ with the minimum Hamming weight is NP-complete~\cite{Berlekamp1978}. Secondly, there are some uncertainties in the message $\mathbf{x}$ especially when the message is compressed. Even an eavesdropper is possible to search all the $2^u$ possibilities of $\mathbf{u}$ with unlimited computing power, given $\mathbf{y}$ and $M$, there are about $2^{O(H(\mathbf{x})+|\mathbf{u}|-|\mathbf{y}|)}$ feasible choices for $\mathbf{x}$. It is almost impossible for an eavesdropper to choose the right one. Thirdly, as in our proposed schemes, the secret key is generated by jointly utilizing all the common secret bits between the two terminals (not block by block). Attacking the system needs to solve the values of the common secret bits together, which is very difficult in a typical application with each node storing more than
gigabytes of secret bits.

\subsection{Network Coding}

According to Theorem \ref{theorem4}, a channel rate $r_{ij}$ reaches its limit if there exists a set of hacked nodes $\mathcal{N}_\textrm{h}$ and a set of channels $P\subseteq \mathcal{P}_\textrm{s}$ that includes the channel $(i,j)$ such that
$$\sum_{(i',j')\in P}r_{i'j'} \geq r_\textrm{secrecy}(P, \mathcal{N}_\textrm{h}).$$
In this case, the channel $(i,j)$ cannot support more communications with end-to-end encryption. But it is still possible for node $i$ to communicate to node $j$ via network coding.

A solution of network coding that can be used here is called ``secret sharing"~\cite{Shamir1979}. In order to tolerate $t$ nodes to be hacked, the source node encodes the message into $t+1$ packets such that no eavesdropper can obtain any information about the message unless getting all the $t+1$ packets. For example, let $\mathbf{x}\in \{0, 1\}^m$ be the message to communicate, then it is encoded into
$$\mathbf{r}_1,\quad \mathbf{r}_2, \quad \ldots, \quad \mathbf{r}_t, \quad \mathbf{r}_{1} +  \ldots + \mathbf{r}_{t}  + \mathbf{x}$$
with the random-bit sequence $\mathbf{r}_i\in \{0, 1\}^m$ as the $i$th packet for $1\leq i\leq t$.  Then the source node sends the $t+1$ packets over node-disjoint paths (only the channels whose rates are below the limit are used) to the destination.  Two nodes can communicate to each other with perfect secrecy if and only if there exits at least $t+1$ node-disjoint paths connecting them.

It is worth mentioning that this network-coding approach based on multiple paths is very expensive: it costs at least $2(t+1)$ times of secret-bit resources (more precisely proportional to the number of channels in the selected $t+1$ node-disjoint paths), and introduces much more communication latency. Furthermore, it may bring in additional adversaries, as some hacked nodes may interrupt the communication by modifying relayed packets or injecting corrupted packets, know as Byzantine adversaries~\cite{Ho2004,Jaggi2007}.

\subsection{Further Questions}

In this paper, we work on a framework that studies the problem of network communication with the information-theoretic security when each node is allowed to carry some pre-distributed randomness. This work is an extension of the well-known one-time pad scheme from `links' to `networks'.  There are several questions that are not completely solved in this paper, which deserve further studies.

\begin{enumerate}
  \item The tradeoff between the maximum network rate and the maximum channel rate for a network without any nodes being hacked is given in Theorem \ref{theorem1}. A natural question is how to extend it to a network with $t>1$.
  \item The criteria in Theorem \ref{theorem4} are both necessary and sufficient for guaranteeing the information theoretic security. It is also proved that the criteria in Theorem \ref{theorem5} are sufficient, but it is unclear whether they are necessary or not.
\end{enumerate}

This paper mainly focuses on networks with symmetric key distribution, where every network node can carry the same number of secret bits. The models, methods and analysis developed can be naturally extended to some other occasions, such as a clustered network or a centralized network. For example, if a network has a trustable central node with a larger storage space, one may distribute all the secret bits to this central node and meanwhile each secret bit is also shared by some of the other nodes. This allows the central node to easily communicate with the other nodes and monitor all the messages transmitted over the network.

\section{Proofs of Main Results}
\label{section_proofs}

In this section we provide proofs of our main results.

\subsection{Proof of Theorem \ref{theorem2}}

The network capacity is easy to derive: The total message length communicated with a node cannot exceed $l$, hence
$$\sum_i (\sum_{j>i} m_{ij} + \sum_{j<i} m_{ji})\leq nl.$$
This leads to $\sum m_{ij} \leq \frac{nl}{2}$, yielding the upper bound $\frac{n}{2}$ on the network capacity. This upper bound is achievable using the simple pairwise key scheme when $t\leq n-2$.
We continue to prove that the channel capacity is at most $\frac{\nchoosek{n-t-2}{a-2}}{\nchoosek{n-1}{a-1}}$ with $a= \lceil\frac{n}{t+1}\rceil$, and it's achievable.

Given the sequence of secret bits stored in node $i$, $\mathbf{u}_i$, for all $i\in \mathcal{N}$, the entropy of $\mathbf{u}_i$ is at most $l$. Due to the symmetry of the network, without loss of generality, we assume that the first $t$ nodes are hacked, then the hacked secret bits are $\mathbf{u}_\textrm{h}=\mathbf{u}_1^t = \mathbf{u}_1 \mathbf{u}_2 \ldots\mathbf{u}_t$, and the maximal number of message bits that can be securely communicated between node $t+1$ and node $t+2$ is the mutual information
between $\mathbf{u}_{t+1}$ and $\mathbf{u}_{t+2}$ conditioning on $\mathbf{u}_1^t$. As a result,
$$\emph{C}_{\textrm{channel}} \leq \frac{\textrm{I}[\mathbf{u}_{t+1}; \mathbf{u}_{t+2}|\mathbf{u}_1^t]}{l}.$$

Since $\textrm{I}[\mathbf{u}_{t+1}; \mathbf{u}_{t+2}|\mathbf{u}_1^t]$ is invariant under any permutation of node indices, for simplicity, we can rewrite it as $\textrm{I}(2|t)$.

We can generalize this concept of conditional mutual information to a higher order as
\begin{align*}
 \textrm{I}(a|b) &= \textrm{I}[\mathbf{u}_{b+1}; \ldots;\mathbf{u}_{b+a}|\mathbf{u}_1^b]  \\
 &= \textrm{I}[\mathbf{u}_{b+2}; \ldots;\mathbf{u}_{b+a}|\mathbf{u}_1^b]  - \textrm{I}[\mathbf{u}_{b+2};\ldots;\mathbf{u}_{b+a}|\mathbf{u}_1^{b+1}],
\end{align*}
with the short notations,
\begin{equation}\textrm{I}(a|b) = \textrm{I}(a-1|b) - \textrm{I}(a-1|b+1).\label{equ3_4}\end{equation}
This quantity $\textrm{I}(a|b)$ is the amount of mutual information among all the $a$ nodes given the secret bits of any other $b$ nodes.

From (\ref{equ3_4}), it can be shown that
\begin{equation}\textrm{I}(1|0) = \sum_{a=1}^{n} \nchoosek{n-1}{a-1} \textrm{I}(a|n-a)\label{equ2_3},\end{equation}
where $\textrm{I}(1|0) = H(\mathbf{u}_i) \leq l $.

On the other hand,
\begin{equation}\textrm{I}(2|t) = \sum_{a=2}^{n-t} \nchoosek{n-t-2}{a-2} \textrm{I}(a|n-a).\label{equ2_5}\end{equation}

From (\ref{equ2_3}) and (\ref{equ2_5}), we get
$$\frac{\textrm{I}(2|t)}{\textrm{I}(1|0)}\leq \max_{a=2}^{n-t} \frac{\nchoosek{n-t-2}{a-2}}{\nchoosek{n-1}{a-1}},$$
where the maximum is achieved with $a= \lceil\frac{n}{t+1}\rceil$. As a result,
$$\textrm{I}(2|t)\leq \frac{\nchoosek{n-t-2}{a-2}}{\nchoosek{n-1}{a-1}}l \quad \textrm{with} \quad a= \lceil\frac{n}{t+1}\rceil. $$
This leads to the upper bound on the channel capacity. This upper bound can be achieved with the combinational key scheme with $a= \lceil\frac{n}{t+1}\rceil$ if the underlying privacy amplification is asymptotically optimal.

\subsection{Proof of Theorem \ref{theorem1}}

Given a scheme $\psi$, let $\mathbf{u}\in \{0, 1\}^u$ be the independent sequence of random bits from which the distributed secret bits are chosen. Denote the fraction of bits that are distributed to exactly $a$ nodes by $p_a$ with $0\leq p_a \leq 1$ and
$$\sum_{a=0} ^n p_a = 1.$$

The total amount of memory needed to store the distributed secret bits is
\begin{equation}\sum_{a=0}^{n}(p_a u) a = (\sum_{a=0}^{n} p_a a)u\leq nl.\label{equ3_5}\end{equation}
Here, we write $\mathbb{E}(f(a)) = \sum_{a=0}^{n}p_a f(a)$ for a function $f(a)$.

Firstly, the total number of secure message bits is upper bounded by the total number of secret bits $u$, hence the maximum network rate
\begin{equation}
\emph{R}_\textrm{net}(\psi) \leq \frac{u}{l} \leq \frac{nl}{\mathbb{E}(a)l}  = \frac{n}{\mathbb{E}(a)}.
\label{equ3_8}
\end{equation}

Secondly, the number of message bits that can be securely communicated over a channel is upper bounded by the number of common secret bits shared by the two terminals. If the key distribution is symmetric, then
\begin{align}\emph{R}_\textrm{channel}(\psi)&\leq \frac{u_{12}}{l} \nonumber\\
&= \frac{1}{l}\sum_{a\geq 2} (u p_a) \frac{\nchoosek{n-2}{a-2}}{\nchoosek{n}{a}} \nonumber \\
&= \frac{u\mathbb{E}(a(a-1))}{l\cdot n(n-1)}\nonumber\\
&\leq \frac{\mathbb{E}(a(a-1))}{\mathbb{E}(a) (n-1)}\nonumber\\
&\leq \frac{n + 1}{n-1}-\frac{2n}{\mathbb{E}(a)(n-1)}. \label{equ3_9}
\end{align}
The last step follows since $a(a-1) \leq (n+1)a - 2n$ for $2\leq a \leq n$.

From(\ref{equ3_8}) and (\ref{equ3_9}), we obtain
$$\emph{R}_\textrm{net}(\psi) \frac{2}{n+1} + \emph{R}_\textrm{channel}(\psi)\frac{n-1}{n+1} \leq 1.$$

Let us prove the achievablity, starting from two simple schemes. In the first scheme (called the same key scheme), all the nodes share the same set of secret bits, and its maximum rates are \begin{equation}\emph{R}_\textrm{net}(\psi_\textrm{same}) = 1, \quad \emph{R}_\textrm{channel}(\psi_\textrm{same}) = 1. \label{equ3_10}\end{equation}

The second scheme is the pairwise key scheme, whose maximum rates are
\begin{equation}\emph{R}_\textrm{net}(\psi_\textrm{pair}) = \frac{n}{2}, \quad \emph{R}_\textrm{channel}(\psi_\textrm{pair}) = \frac{1}{n-1}. \label{equ3_11}\end{equation}

The equality in the theorem holds both for the same key scheme and the pairwise key scheme. Here we construct a scheme as the hybrid of the two simple schemes. For each node,
it uses a fraction $0\leq \lambda \leq 1$ of its storage space for the same key scheme and the rest for the pairwise key scheme. The maximum rates for the hybrid scheme $\psi_\textrm{hybrid}$  are
\begin{align*}
\emph{R}_\textrm{net}(\psi_\textrm{hybrid}) & = \lambda + \frac{n}{2}(1-\lambda),\\
\emph{R}_\textrm{channel}(\psi_\textrm{hybrid}) & = \lambda + \frac{1}{n-1}(1-\lambda).
\end{align*}
By adjusting the fraction $\lambda$, we can obtain all the maximum network rates and the maximum channel rates meeting the equality in the theorem.

\subsection{Proof of Theorem \ref{theorem4}}

The necessity is easy to prove: if there exists a subset of channels violating (\ref{equ3_3}), then their total message length must be larger than the number of their used unhacked secret bits. As a result, at least one of these messages must be information-theoretically insecure.

To prove achievability, we consider a simple method for privacy amplification: for every channel $(i, j)$, given the common secret bits $\mathbf{u}_{ij}$, the secret key $\mathbf{s}_{ij}=M_{ij}\mathbf{u}_{ij}$ with a sparse random matrix $M_{ij}$ of density $O(\log l/l)$. The reason of using this method is not only due to its asymptotic optimality,
but also to its practicality. It is the basis of our proposed network schemes.

Let $\mathbf{s}_\textrm{s}= \{\mathbf{s}_{ij}|(i, j) \in \mathcal{P}_\textrm{s}\}$ be the secret keys between unhacked nodes, and let $\mathbf{u}_\textrm{h}$  be the distinct secret bits stored in hacked nodes, which are disclosed to the eavesdropper. The network communication is information-theoretically secure if and only if for any possible set of hacked nodes $\mathcal{N}_{h}$, the secret keys $\mathbf{s}_\textrm{s}$ and the hacked secret bits $\mathbf{u}_\textrm{h}$ are truly random bits, and $\mathbf{s}_\textrm{s}, \mathbf{u}_\textrm{h}$ are independent.
Note that both $\mathbf{s}_\textrm{s}$ and $\mathbf{u}_\textrm{h}$ can be written as linear transformations of the source sequence $\mathbf{u}$.

Let $\mathbf{z}$ be the concatenation of $\mathbf{s}_\textrm{s}$ and $\mathbf{u}_\textrm{h}$, then
\begin{equation}\mathbf{z}= \mathbf{s}_\textrm{s}\mathbf{u}_\textrm{h} = M \mathbf{u}\label{equ3_1}\end{equation}
for some matrix $M$. The network is information-theoretically secure if and only if all the rows in matrix $M$ are linearly independent.

We can write the secret key $\mathbf{s}_{ij}$ as
\begin{align*}
   \mathbf{s}_{ij} & = A_{ij}^{'}(\mathbf{u}_{ij}/\mathbf{u}_\textrm{h}) + B_{ij}^{'} (\mathbf{u}_{ij}\cap\mathbf{u}_\textrm{h})\\
    &= A_{ij} (\mathbf{u}/\mathbf{u}_\textrm{h}) + B_{ij} \mathbf{u}_h
\end{align*}
for some matrices $A_{ij}$ and $B_{ij}$, where $A_{ij}$ is an $(lr_{ij})\times |\mathbf{u}/\mathbf{u}_\textrm{h}|$ matrix consisting of $|\mathbf{u}_{ij}/\mathbf{u}_\textrm{h}|$
random columns of density $O(\log l /l )$ and $|\mathbf{u}/\mathbf{u}_\textrm{h}| - |\mathbf{u}_{ij}/\mathbf{u}_\textrm{h}|$ zero columns.

Then $\mathbf{z}= \mathbf{s}_\textrm{s} \mathbf{u}_\textrm{h}$ is represented by
\begin{equation} \mathbf{z} =  M\mathbf{u} = \left(
                  \begin{array}{cc}
                    A & B \\
                    0 & I \\
                  \end{array}
                \right) \left(
                          \begin{array}{c}
                             \mathbf{u}/\mathbf{u}_\textrm{h} \\
                            \mathbf{u}_\textrm{h} \\
                          \end{array}
                        \right),\label{equ3_6}
\end{equation}
where $I$ is an identity matrix and $A$ consists of all the matrices $A_{ij}$ with $(i, j)\in \mathcal{P}_\textrm{s}$. The network is perfectly secure if the rows in $M$ are linearly independent. This is equivalent to showing that the rows in $A$ are linearly independent, i.e., all the rows in $\{A_{ij}|(i,j)\in \mathcal{P}_\textrm{s}\}$ are linearly independent. This can be proved based on the following results.

\begin{Lemma} All the rows in $\{A_{ij}|(i,j)\in \mathcal{P}_\textrm{s}\}$ are linearly independent if and only if for any subset of channels $P\subseteq \mathcal{P}_\textrm{s}$,
there does not exist any subset of rows from $\{A_{ij}|(i,j)\in P\}$ that includes at least one row from each matrix such that their sum is a zero-vector.
\end{Lemma}

\begin{Lemma} \label{lemma1}Given any subset of channels $P\subseteq \mathcal{P}_\textrm{s}$, if $\mathbf{s}_{ij}=M_{ij}\mathbf{u}_{ij}$ with a random matrix $M_{ij}$ of density $O(\log l/l)$ and
$$\frac{\sum_{(i,j)\in P} |\mathbf{s}_{ij}|}{|\cup_{(i,j)\in P} \mathbf{u}_{ij} / \mathbf{u}_\textrm{h}|} < 1$$
with $|\cup_{(i,j)\in P} \mathbf{u}_{ij} / \mathbf{u}_\textrm{h}|=O(l)$, when $l\rightarrow\infty$, with probability almost $1$ there does not exist any subset of rows from $\{A_{ij}|(i,j)\in P\}$ that includes at least one row from each matrix such that their sum is a zero-vector.
\end{Lemma}

The proof of Lemma \ref{lemma1} is provided in subsection \ref{subsection_lemma1}. Finally, we can conclude that the rows of the security matrix $M$ are linearly independent with high probability, and the criteria in Theorem \ref{theorem4} are sufficient.

\subsection{Proof of Theorem \ref{theorem5}}

Using the same proof as Theorem \ref{theorem4}, the network is perfectly secure if and only if the rows of the matrix $A$ in (\ref{equ3_6}) are linearly independent.
In Theorem \ref{theorem5}, for this matrix $A$, it has the following properties: there are $|\mathbf{u}_G/\mathbf{u}_\textrm{h}|$ columns in $A$ corresponding to the bits in $\mathbf{u}_G/\mathbf{u}_\textrm{h}$, in which
each column has $\sum_{(i,j)\in G} m_{ij}$ random entries with $m_{ij}=l\cdot r_{ij}$ corresponding to the bits in $\{\mathbf{s}_{ij}\}$ with $i, j \in G$.

The rank of the matrix $A$ remains unchanged if we do elementary row or column operations on $A$. The rows of a matrix are linearly independent if and only if the
the matrix can be reduced to the simplest form $[I, 0]$ by elementary operations such that it consists of an identity matrix and a zero matrix.

If there exists a feasible solution for $\{x_{ij}^G\}$, we can divide the columns corresponding to the bits in $\mathbf{u}_G/\mathbf{u}_\textrm{h}$ into some groups of sizes $\{u_{ij}^{G}\}$ with $u_{ij}^G = l\cdot x_{ij}^G$ and $\sum_{i,j\in G} u_{ij}^G = |\mathbf{u}_G/\mathbf{u}_\textrm{h}|$.

On the other hand, we can divide the rows corresponding to the bits in $\mathbf{s}_{ij}$ into some groups of sizes $\{m_{ij}^{G}\}$ with $m_{ij}^G = l \cdot y_{ij}^G$ and
$$y_{ij}^G = \frac{r_{ij}}{\sum_{G}x_{ij}^G}x_{ij}^G, \sum_{G|i,j\in G} m_{ij}^G= |\mathbf{s}_{ij}|.$$
According to the inequalities in the theorem, it has either $y_{ij}^G<x_{ij}^G$ or $y_{ij}^G=0$.

Based on the row groups and the column groups, the matrix $A$ is divided into $|\{m_{ij}^G\}|\times |\{u_{ij}^G\}|$ sub-matrices, whose dimensions are
$\{m_{ij}^{G}\}\times \{u_{ij}^{G}\}$. By switching the rows and columns of the matrix $A$, the matrix $A$ can be transformed into a form such that the sub-matrices of dimensions
$\{m_{ij}^{G} \times u_{ij}^{G}\}$ are on the diagonal of the sub-matrices. We denote the sub-matrices on the diagonal by $[A_1, A_2, \ldots] = \{A_{ij}^G\}$, and the matrix $A$ is transformed to
$$A\Leftrightarrow\left(
    \begin{array}{ccc}
      A_1 & \ldots & \ldots \\
      \vdots & A_2 & \vdots \\
      \vdots & \ldots & \ddots  \\
    \end{array}
  \right).
$$
The sub-matrices $[A_1, A_2, \ldots]$ are random matrices of density $O(\log l/l)$. The dimension of the sub-matrix $A_i$ is $m_i \times u_i$ for some $m_i, u_i$ such that $\frac{m_i}{u_i} < 1$ for $u_i = O(l)$ or $m_i=0$.

For the sub-matrix $A_1$, according to Lemma \ref{lemma2} in subsection \ref{subsection_lemma1}, the rows of $A_{1}$ are linearly independent with high probability when $l$ is sufficiently large. The sub-matrix $A_1$ can be reduced to its simplest form $[I_1, 0]$ consisting of an identity matrix and a zero matrix by elementary operations on $A$. Furthermore, all the other entries on the right of $A_1$ (in the same rows with $A_1$) can be reduced to $0$ by elementary column operations. Right now, each sub-matrix $A_i$ with $i>1$ is transformed to $A_i'$ with
$$A_i' = A_i + \overline{A_i}$$
for some $\overline{A_i}$ independent of $A_i$, and the matrix $A$ is reduced to
$$A\Leftrightarrow\left(
    \begin{array}{ccc}
      I_10 & 0 & 0 \\
      \vdots & A_2' & \vdots \\
      \vdots & \ldots & \ddots  \\
    \end{array}
  \right).
$$

We continue repeating the above process to handle $A_2', A_3', ..., $ iteratively. For the sub-matrix $A_i'=A_i + \overline{A_i}$, it can be proved that the conclusion of Lemma \ref{lemma2} still holds, and all the rows of  $A_i'$ are linearly independent with high probability when $l$ is sufficiently large.

Finally, all the sub-matrices $[A_1, A_2, ...]$ are reduced to their simplest forms with high probability, and all the other entries on their right are $0$s. In this case, the matrix $A$ is reduced to the reversed row echelon form, and it has full rank. Hence all the rows of the matrix $A$ are linearly independent with high probability if $l$ is sufficiently large. This leads to the achievability of the channel rates.

\subsection{Proof of Lemma \ref{lemma1}}
\label{subsection_lemma1}

We first prove the following result.

\begin{Lemma}\label{lemma2}
Let $M\in \{0,1\}^{k\times r}$ be a random matrix such that the probability of each entry being $1$ is $p=O(\log r/r)$. The rows in $M$ are linearly independent with high probability for sufficiently large $r$ if and only if $k/r<1$.
\end{Lemma}

Each row in $M$ is an independent random vector. The sum of any $j$ rows in $M$ is still an independent vector. Denote the probability of its entry being $1$ by $p_j$.

It is easy to show that
$$p_{j} = p_{j-1}(1-p)+ (1-p_{j-1})p,$$
from which and by induction, we obtain
$$p_j = \frac{1}{2}-\frac{1}{2}(1-\frac{2d}{r})^j.$$

Furthermore, since the sum of any $j$ rows is an independent vector, the probability for it being a zero-vector is
$$\mathbb{P}_j(0) = (1-p_j)^r.$$

The rows of $M$ are linearly independent if and only if for any subset of the rows, their sum is not a zero-vector. Hence, the probability of the rows of $M$ being linearly independent
$$ \mathbb{P}_{\mathrm{indep}}(M) \geq 1- \sum_{j=1}^k \nchoosek{k}{j}\mathbb{P}_j(0),$$
where $\nchoosek{k}{j}$ is the number of subsets consisting of $j$ rows.

This leads to
\begin{equation}\mathbb{P}_{\mathrm{indep}}(M)\geq 1- \sum_{j=1}^k \nchoosek{k}{j} (\frac{1}{2}+\frac{1}{2}(1-\frac{2d}{r})^j)^r.\label{equ2_11}\end{equation}

When $j<\frac{r}{2d}$ with $r$ sufficiently large, it has
\begin{align*}
    &\sum_{j=1}^{\frac{r}{2d}}\nchoosek{k}{j} (\frac{1}{2}+\frac{1}{2}(1-\frac{2d}{r})^j)^r\\
    \leq &  \sum_{j=1}^{\frac{r}{2d}}k^j (1 - \frac{dj}{r} + \frac{j(j-1)d^2}{r^2})^r\\
    \leq &  \sum_{j=1}^{\frac{r}{2d}}k^j (1 - \frac{dj}{2r})^r\\
    \leq & \sum_{j=1}^{\frac{r}{2d}}  r ^j e ^{-dj/2}\\
    \leq & \sum_{j=1}^{\frac{r}{2d}}   (e ^{\log r-d/2})^j \rightarrow 0.
\end{align*}

When $\frac{r}{2d}\leq j < \frac{r}{3\log r}$ with $r$ sufficiently large, it has
\begin{align*}
    &\sum_{j=\frac{r}{2d}}^{ \frac{r}{3\log r}}\nchoosek{k}{j} (\frac{1}{2}+\frac{1}{2}(1-\frac{2d}{r})^j)^r\\
    \leq &  \sum_{j=\frac{r}{2d}}^{ \frac{r}{3\log r}}\nchoosek{k}{j} (\frac{1}{2}+\frac{e^{-1}}{2})^r \\
    \leq &  r ^{ \frac{r}{3\log r}+1}(\frac{1}{2}+\frac{e^{-1}}{2})^r  \\
    \leq & r (e^{\frac{1}{3}} (\frac{1}{2}+\frac{e^{-1}}{2}))^r\rightarrow 0.
\end{align*}

When $j\geq \frac{r}{3\log r}$ with $r$ sufficiently large, it has
\begin{align*}
    &\sum_{j=\frac{r}{3\log r}}^{ k}\nchoosek{k}{j} (\frac{1}{2}+\frac{1}{2}(1-\frac{2d}{r})^j)^r\\
    \leq &  \sum_{j=\frac{r}{3\log r}}^{ k}\nchoosek{k}{j} (\frac{1}{2}+\frac{e^{-\frac{2d}{3\log r}}}{2})^r \\
    \leq & 2 ^k  (\frac{1}{2}+\frac{e^{-\frac{2d}{3\log r}}}{2})^r \\
    = & (2^{\frac{k}{r}-1}(1+e^{-\frac{2d}{3\log r}}) )^r\\
    \rightarrow & (2^{\frac{k}{r}-1})^r\rightarrow 0.
\end{align*}

Summing the above results up, we obtain
$$\mathbb{P}_{\mathrm{indep}}(M)\geq 1-\epsilon$$
for any $\epsilon>0$ when $r$ is sufficiently large.

\begin{Lemma}\label{lemma3}
Given $S_i\subseteq\{1, 2, \ldots, r\}$ with $1\leq i \leq k$, let $M_i\in \{0, 1\}^{m_i\times r}$ with $1\leq i\leq k$ be a binary matrix such that each entry in columns $S_i$ is $1$ with probability $O(\log r/r)$ and each entry not in columns $S_i$ is $0$. If $\frac{\sum_i m_i}{|\cup_i S_i|} < 1$ and  $|\cup_i S_i| =O(r)$, as $r\rightarrow \infty$,  with high probability there does not exist any subset of rows from $\{M_i\}$ that includes at least one row from each matrix such that their sum is a zero-vector.
\end{Lemma}

We say that a set of matrices $\{M_i\}$ are linearly cross-independent if and only if  there does not exist a subset of rows from $\{M_i\}$ that includes at least one row from each matrix such that their sum is a zero-vector. If the rows of $M_1, M_2, \ldots, M_k$ are linearly cross-independent, it does not necessarily imply that these rows are linearly independent. For example, consider the matrices
$$M_1 = \left(
          \begin{array}{ccc}
            1 & 1 & 0 \\
            1 & 1 & 0 \\
           \end{array}
        \right), M_2 = \left(
          \begin{array}{ccc}
            0 & 1 & 1 \\
           \end{array}
        \right)
$$
The rows in $M_1, M_2$ are linearly cross-independent, but not linearly independent, as the rows in $M_1$ are not linearly independent.

One observation is that if $\{M_i\}$ are linearly cross-independent on a subset of columns, then $\{M_i\}$ are linearly cross-independent on all the columns.

We divide the columns into at most $2^k$ groups depending on which $S_i$ the column belongs to. Two columns are in the same group if and only if they belong to the same subset of $\{S_1, S_2, \ldots, S_k\}$. Now, we are only interested in the groups of size $O(r)$ (sufficiently large groups), and the union of their columns are denoted by $S$. Then
$$|S|> |\cup_i S_i| (1-\epsilon) $$
for sufficiently small $\epsilon$, which leads to $\frac{\sum_i m_i}{|S|} < 1$ for sufficiently large $r$.
We will prove that the matrices $\{M_i\}$ are linearly cross-independent on the columns in $S$.

Given $\ell= \{l_1, \ldots, l_k\}$ with $1\leq l_i \leq m_i$, we choose $l_i$ rows from $M_i$ with $1\leq i \leq k$, and we use $P(\ell)$ to denote the probability that
the sum of all the $\sum_i l_i$ chosen rows is a zero-vector on $S$. Then the probability of the matrices $\{M_i\}$ being not linearly cross-independent on $S$ is
$$\mathbb{P}_\textrm{dep.} = \sum_{l_1, l_2, \ldots} \prod_{i=1}^{k}\nchoosek{m_i}{l_i}P(\ell).$$

There are two possibilities considering the chosen $\sum_i$ rows: (1) every column in $S$ has more than $\frac{r}{\log r}$ random entries in the chosen rows; and (2) there exists a group  (among the up to $2^k$ groups)  of columns in $S$, whose size is at least $b=O(r)$ and in which each column has at most  $\frac{r}{\log r}$ random entries in the chosen rows. We use $P_1(\ell)$ to denote the probability that the sum of chosen rows is a zero-vector on $S$ in the first case, and $P_2(\ell)$ to denote that in the second case. It can be shown that
$$P_1(\ell) \leq (\frac{1}{2}+\epsilon)^{|S|}$$
for sufficient small $\epsilon$ and
$$P_2(\ell) \leq 2^k \sum_{j=1}^{\frac{r}{\log r}} \nchoosek{\sum_i l_i}{j} (\frac{1}{2}+\frac{1}{2}(1-\frac{2d}{r})^j)^b.$$

Consider all possible choices of $\ell$, as $r \rightarrow \infty$, the sum probability of the first case is
\begin{align*}
   \mathbb{P}_{1}&= \sum_{l_1, l_2, \ldots} \prod_{i=1}^{k}\nchoosek{m_i}{l_i}P_1(\ell)\\
   & \leq \sum_{l_1, l_2, \ldots} \prod_{i=1}^{k}\nchoosek{m_i}{l_i}(\frac{1}{2}+\epsilon)^{|S|}\\
    &\leq 2^{\sum_{i} m_i}(\frac{1}{2}+\epsilon)^{|S|}\\
    &\leq \epsilon
\end{align*}
for sufficiently small $\epsilon$.

Consider all possible choices of $\ell$, as $r \rightarrow \infty$, the sum probability of the second case is
\begin{align*}
\mathbb{P}_{2}&= \sum_{l_1, l_2, \ldots} \prod_{i=1}^{k}\nchoosek{m_i}{l_i}P_2(\ell)\\
&\leq \sum_{j=1}^{\frac{r}{\log r}} \nchoosek{\sum_i m_i}{j} (\frac{1}{2}+\frac{1}{2}(1-\frac{2d}{r})^j)^b\\
&\leq \epsilon
\end{align*}
for sufficiently small $\epsilon$. The last step follows the same proof as Lemma \ref{lemma2}.

Finally, the matrices $\{M_i\}$ are not linearly cross-independent with probability
$$\mathbb{P}_\textrm{dep.}\leq \mathbb{P}_{1} + \mathbb{P}_{2} \leq  2\epsilon$$
as $r\rightarrow \infty$.  This leads to the conclusion in Lemma \ref{lemma3}.

It is straightforward to obtain Lemma \ref{lemma1} from Lemma \ref{lemma3}.

\ifCLASSOPTIONcaptionsoff
  \newpage
\fi

\end{document}